\documentclass[prl,twocolumn]{revtex4}
\usepackage{amssymb}
\usepackage{graphicx}
\usepackage{amsfonts}
\usepackage{amsmath}

\begin{document}

\title{Electrically Controlled  Pumping of Spin Currents in Topological Insulators}

\author{R. Citro and F. Romeo}
\affiliation{Dipartimento di Fisica ``E. R. Caianiello'',
Universit{\`a} degli Studi di Salerno, and Institute CNR-SPIN, UO Salerno,  Via Ponte don Melillo,
I-84084 Fisciano (Sa), Italy}
\author{N. Andrei}
\affiliation{Center for Materials Theory, Serin Physics Laboratory, Rutgers University, Piscataway, New Jersey 08854-8019, USA}

\date{\today}

\begin{abstract}
Pure spin currents are shown to be generated by an \emph{electrically controlled} quantum pump applied at the edges of a topological insulator. The electric rather than the more conventional magnetic control offers several advantages and avoids, in particular, the necessity of delicate control of magnetization dynamics over tiny regions. The pump is implemented by pinching the sample at two quantum point contacts and phase modulating two external gate voltages between them. The spin current is generated for the full range of parameters.
On the other hand, pumping  via amplitude modulation of the inter-boundary couplings generates both charge and spin currents, with a pure charge current appearing only for special values of the parameters for which the Bohm-Aharonov flux takes integer values.
Our setup can therefore serve to fingerprint the helical nature of the edges states with the zeros of the pumped spin and charge currents occurring at distinct universal locations where the
Fabry-P\'{e}rot or the Aharonov-Bohm phases take integer values.
\end{abstract}



\maketitle

Quantum transport phenomena are fundamental  in many areas of physics as well as in chemistry, and biology.
An intriguing example is the quantum pumping, where, in the absence of external bias,
a directed current of particles is produced along a periodic
structure\cite{thouless_qpt,alts_qpt}  by a slow periodic variation of some
system characteristic; the variation being slow enough so
that the system remains close to its ground state throughout
the pumping cycle. From a fundamental physics standpoint,
this mechanism represents a new macroscopic quantum
phenomenon reminiscent of the quantum Hall effect and of
superconductivity where current flows without dissipation.

Experiments aimed at observing this phenomenon have been carried out in various setups:  charge pumping was carried out in semiconductor quantum
dots\cite{switkes_qp_dot}, quantum wires\cite{qp_nanowire1,qp_nanowire2} and also in carbon nanotubes\cite{qp_nanotube}, while
 spin pumping, important in spintronics applications\cite{bauer_physics}, was mainly proposed in quantum wires and quantum rings\cite{spin-pump} and realized in GaAs based
 quantum dots\cite{spin_pump_exp_dot} and very recently in insulating ferromagnets\cite{sandweg_exp_magnon_pump}.

The recent discovery of Topological insulators (TI), bulk gapped materials exhibiting conducting channels at the boundaries\cite{various_ti} represents a new promising route to spin manipulation in semiconductors.
In the presence of edges states the generation of spin currents is naturally facilitated by their helical nature: namely, in a two-dimensional realization of a TI, only spin-up electrons propagate rightwards and only spin-down electrons
leftwards, along a given boundary. These  helical edge states behave as perfectly conducting one-dimensional channels, in which backscattering off an impurity is prevented by time-reversal symmetry. Following the theoretical predictions\cite{kane_ti_2005,zhang_ti_2006},
a successful realization of a 2D TI phase has been obtained in HgTe/CdTe quantum wells (QWs)\cite{konig_ti_2007,konig_ti_2008}
making TI ideal candidates for spintronics devices.\cite{various_ti,spintronics-ti-1,spintronics-ti-2,spintronics-ti-3}

Natural questions raised by these rapid developments in the field
are: how can setups exploit TI
edge or surface states in the presence of external magnetic fields used to
manipulate spins or of electrical fields not breaking the time-reversal invariance?
How can the 2D TI phase be detected by conventional measurements of charge transport quantities?

\begin{figure}[h]
\centering
\includegraphics[clip,scale=0.5]{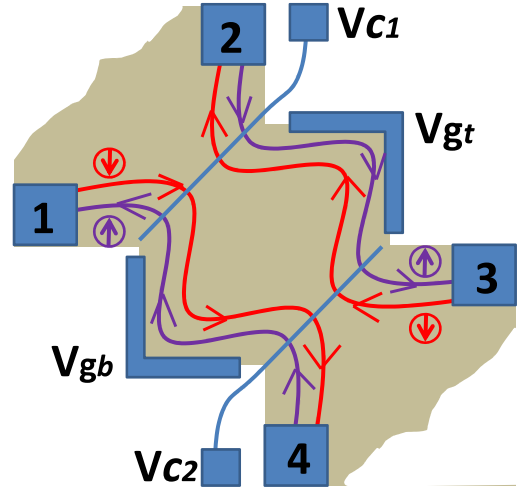}\\
\caption{Representation of the edge state flow at the top and bottom boundaries of a TI quantum well. The straight lines indicate the scattering region, where the inter-boundaries tunneling events occur at the two QPCs at varying the gate voltages $V_{ci}$. The boxes illustrate the four-terminal device.}
\label{fig:device}
\end{figure}

In this Letter we aim at answering the above questions by proposing a quantum pumping four terminal setup in which the interedges coupling can be controlled at two quantum point contacts either by amplitude modulation or by phase modulation pumping by {\it all electrical} means, i.e. harmonically varying external gate voltages as shown in Fig.\ref{fig:device}. A similar setup was considered in the context of electron interferometry in Ref.[\onlinecite{dolcini_ti_archive}]. Other spin pumping mechanisms in 2D TI utilizing time-dependent magnetic fields\cite{brouwer_2010,zhang_nature_2008} or magnetization dynamics\cite{chang_2011} have recently appeared. Our  proposal, we believe, is simpler to implement as it does not require  nanomagnet contacts and  magnetization dynamics control.

We consider a two-dimensional realization of a TI where Kramers pairs edge states flow at the top and bottom boundaries of a quantum well. Inter-boundary scattering can occur at two quantum point contacts (QPCs), giving rise to loop trajectories. The tunneling terms can be modulated in time either by point-like voltages $V_{C1}$ and  $V_{C2}$ located at $x_1$ and $x_2$  at a distance $L$ apart, or by top and bottom gate voltages $V_{gt},V_{gb}$ whose effect is to modify the electron phase in the loop paths by shifting the electron momenta in the region between the two QPCs.
We now show that the setup of Fig.\ref{fig:device} can lead to a fully electrical controllable system where pure charge or pure spin current can be generated by the slow periodic variation of two out-of-phase gate voltages.

The edges states at the boundaries of the device are described, at low energies, by the following Hamiltonian\cite{chamon_ti_2009}:
\begin{eqnarray}
H_0=-i\hbar v_F \sum_{\sigma=\uparrow,\downarrow}\int dx && \lbrack : \psi^\dagger_{R\sigma}(x)\partial_x \psi_{R\sigma}(x): \nonumber \\
&&  -: \psi^\dagger_{L\bar{\sigma}}(x)\partial_x \psi_{L\bar{\sigma}}(x):\rbrack
\end{eqnarray}
where $\psi_{R(L)\sigma}$ is the right (left) mover electron field with spin $\sigma=\uparrow,\downarrow$ and $: :$ stands for the normal ordering with respect to the equilibrium state where all levels below Fermi level are occupied. In our description we assume that spin-$\uparrow$ right movers $(R, \uparrow)$ and spin-$\downarrow$ left movers  $(L, \downarrow)$  flow along the top boundary while the spin-$\downarrow$ right movers $(R, \downarrow)$ and spin-$\uparrow$ left movers $(L, \uparrow)$ flow along the bottom boundary. This basis is particularly suitable to study the current at different terminals.\\
The presence of two QPCs at the positions $x_1$ and $x_2$ induces inter-boundary tunneling events. The only tunneling terms which preserve time-reversal symmetry\cite{zhang_ti_2006} can be distinguished in a spin-preserving tunneling
\begin{eqnarray}
\label{eq:backscattering}
H_{sp}=\sum_{\sigma=\uparrow,\downarrow} \int dx &&\lbrack \gamma_{sp}(x,t) \psi^\dagger_{R\sigma}(x)\psi_{L\sigma}(x)+\nonumber \\
&&+\gamma_{sp}(x,t)^\star \psi^\dagger_{L\bar{\sigma}}(x)\psi_{R\bar{\sigma}}(x)
\rbrack
\end{eqnarray}
and a spin-flipping tunneling:
\begin{eqnarray}
\label{eq:spin-flip}
H_{sf}=\sum_{\alpha=L,R} \int dx &&\xi_\alpha \lbrack \gamma_{sf}(x,t) \psi^\dagger_{\alpha\uparrow}(x)\psi_{\alpha\downarrow}(x)+\nonumber \\
&&+\gamma_{sf}(x,t)^\star \psi^\dagger_{\alpha\downarrow}(x)\psi_{\alpha\uparrow}(x)
\rbrack,
\end{eqnarray}
where $\alpha=L,R$ and $\xi_R=+1,\xi_L=-1$ is the chirality. The last tunneling term arises from the local modification of the spin-orbit coupling due to the constriction with respect to the bulk case\cite{spin-orbit-tunnel}. The $\gamma_{i}(x,t)$, ($i=sp,sf$) denote the space and time dependent tunneling amplitudes.
Two external gate voltages, $V_{c1}$ and $V_{c2}$, are used to control the amplitude and the time variation of the tunneling parameters $\gamma_i$ in the Hamiltonian while their space profile is determined by the spatial constriction which is peaked around the two centers $x_1$ and $x_2$ and rapidly decaying beyond a longitudinal length scale $l$. They are assumed of the form:$\gamma_i(x,t)=\sum_{m=1,2} \gamma_{i,m}(x,t)= \sum_{m=1,2}\lbrack \gamma_{i,m}^0(x)+\gamma_{i,m}^\omega(x)\cos (\omega_0t+\varphi_m)\rbrack$, where $m$ denotes the constriction. In the case in which the distance $L=x_2-x_1$ between the two QPCs is large compared to the length scale $l$ and if the constrictions are short compared to the Fermi wavelengths $\lambda_F$, i.e. $l \le \lambda_F < L$, it is sufficient to assume that the tunneling is point-like, i.e. $\gamma_{i,m}^{0,\omega}(x)=2\hbar v_F  \tilde{\gamma}_{i}^{0,\omega}  \delta (x-x_m)$ where $\tilde{\gamma}_{i}$ are dimensionless tunneling amplitudes. Finally, the coupling of top and bottom boundaries to the external gates $V_{gt}$ and $V_{gb}$  is described by the Hamiltonian:
\begin{eqnarray}
\label{eq:gates}
H_g=\int_{x_1}^{x_2} dx \lbrack e V_{gt}(t) (\rho_{R \uparrow}+\rho_{L\downarrow})+ \nonumber \\
e V_{gb}(t)(\rho_{R \downarrow}+\rho_{L\uparrow}) \rbrack,
\end{eqnarray}
where $\rho_{\alpha \sigma}=:\psi^\dagger_{\alpha \sigma}\psi_{\alpha \sigma} :$ denotes the electron density with $\alpha=L,R$ and spin $\sigma$. The gate voltage is varied in time as: $V_r(t)=V_r^0+V_r^\omega \cos(\omega_0 t+\varphi_r)$, $r=gt,gb$,  where in the weak pumping regime $V_r^\omega \ll V_r^0$. The presence of such voltages breaks the degeneracy between top and bottom boundaries by linearly shifting the electron momenta giving a different phase for the electron wavefunctions traveling rightwards and leftwards along the loop created by the two QPCs. In fact, the electron phase accumulated in the loop is $\pi \phi_{FP}=e(V_{gt}+V_{gb})L/\hbar v_F$ in the case of spin-preserving processes ($\gamma_{sp}\ne 0,\gamma_{sf}=0$) and $\pi \phi_{AB}=e(V_{gt}-V_{gb})L/\hbar v_F$ for spin-flipping tunneling ($\gamma_{sp}= 0,\gamma_{sf}\ne 0$). The loop processes induced by the tunneling events  are reminiscent of the Fabry-P\'{e}rot (FP) interference and Aharonov-Bohm (AB) interference phenomena,
respectively\cite{dolcini_ti_archive}.\\

An amplitude modulating pumping (AMP) is realized by harmonically varying the QPCs external voltages $V_{c1}$ and $V_{c2}$ with a frequency $\omega_0$ which modifies the time dependent tunneling amplitude profiles $\gamma_i(x,t)$ as described above, while the top and bottom gates are kept constant. The phase modulating pumping (PMP) can instead be realized by harmonically varying the top (T) and bottom (B) gates, while keeping $V_{c1},V_{c2}$ constant. The variation of $V_{gt,gb}$ induces a time dependence of the electron phase along the top and bottom boundaries\cite{note1} of the form $\phi_{\beta}(t)=\phi_{\beta}+\phi^{\omega}_{\beta}\cos(\omega_0t+\varphi_{\beta})$, with $\beta=T$ or $B$.

The adiabatic pumped charge current per channel $i\equiv (\alpha,\sigma)$\cite{note2} can be obtained by the parametric derivatives of the scattering matrix connecting incoming and outgoing electrons field operators\cite{buttiker_92} at the two QPCs as:
\begin{eqnarray}
J^{pump}_i=\frac{e\omega_0}{2\pi} \eta_i (-)^{i} X_1^\omega X_2^\omega \sin \varphi \\\nonumber
\sum_{j} \text{Im} \{ (\partial_{X_1} S_{ij}^\star)(\partial_{X_2} S_{ij})\},
\end{eqnarray}
where $X_{1,2}^\omega$ denotes the amplitude of time-varying parameter, either $\gamma_{i}$ or $V_r$ and $\eta_i=\pm 1$ is the helicity for edge states belonging to the top/bottom boundary and $\varphi$ the phase-shift between the two parameters. $S_{ij}$ is a four by four matrix whose diagonal entries vanish by helicity and time-reversal symmetry while all the other entries
can be explicitly determined as a function of the tunneling amplitudes $\gamma_{sp},\gamma_{sf}$ via imposing the proper boundary conditions on the wave functions and current conservation at the QPCs. The explicit expressions for the entries of the $S_{ij}$ of a single QPC are: $S_{12}=S_{21}=S_{34}=S_{43}=-2 i \gamma_{sp}/(1+\gamma_{sp}^2+\gamma_{sf}^2)$, $S_{13}=-S_{31}=S_{24}=S_{42}=2 i \gamma_{sf}/(1+\gamma_{sp}^2+\gamma_{sf}^2)$, $S_{14}=S_{41}=-S_{23}=S_{32}=(\gamma_{sp}^2+\gamma_{sf}^2-1)/(1+\gamma_{sp}^2+\gamma_{sf}^2)$, while the total scattering matrix is obtained by composition rule\cite{composition-scattering-matrix}.
Let us note that the parametric derivatives with respect to the gate voltages $V_{gt,gb}$ can be conveniently expressed in terms of the derivatives with respect to the FP and AB phases.\\
Typically,
one has a pumped spin current in addition to the charge current: $I_s^{pump}=\sum_{\sigma=\uparrow,\downarrow} \sigma (I_{R\sigma}^{pump}-I_{L \sigma}^{pump})$  because electron
trajectories that enclose the same geometrical area (see Fig.\ref{fig:device}) pick-up opposite sign  AB fluxes  for electrons injected in the top or bottom boundary. The helical properties of the TI edge states relate
the AB phase $\phi_{AB}$ to spin-flipping phenomena, thus modifying the spin-current. In the case of the AMP, the response to the parametric variation can be interpreted as a backscattering current in the charge sector for the spin-conserving tunneling processes $I_c=i[\rho_{L\sigma},H_b]=-i[\rho_{R\sigma},H_b]$ and a similar contribution in the spin sector coming from the spin-flipping tunneling events $I_s=i[\rho_{\alpha\uparrow},H_s]=-i[\rho_{\alpha\downarrow},H_s]$. The adiabatic pumped current can be calculated as the linear response to the time-dependent Hamiltonian $\delta H(t)=H_{sp}(t)+H_{sf}(t)$: $I^{pump}\simeq i\sum_\beta \int_{-\infty}^t dt' \langle[I_\beta(t),\delta H(t')]\rangle_0$ ($\beta=c,s$), yielding the same result of the scattering matrix approach when considering only terms of second order in the tunneling.

\begin{figure}[h]
\centering
\includegraphics[scale=0.6]{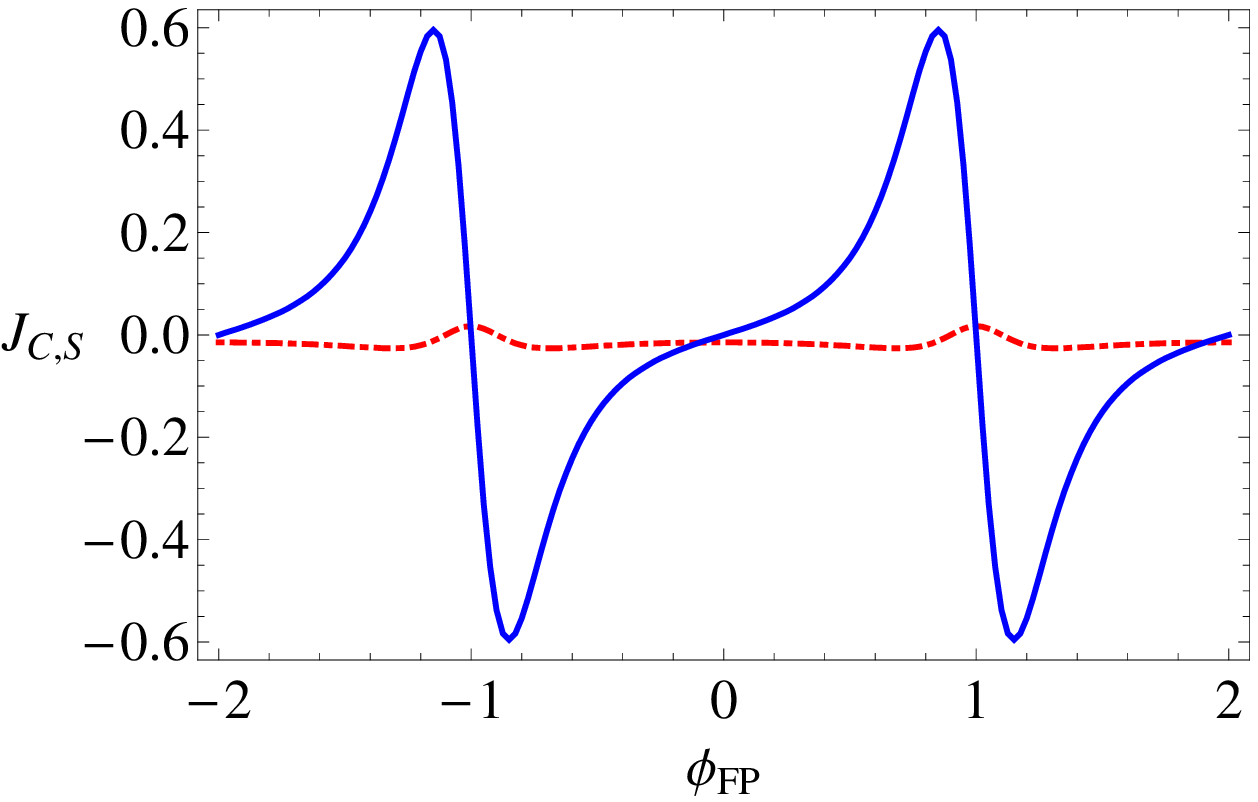}\\
\includegraphics[scale=0.6]{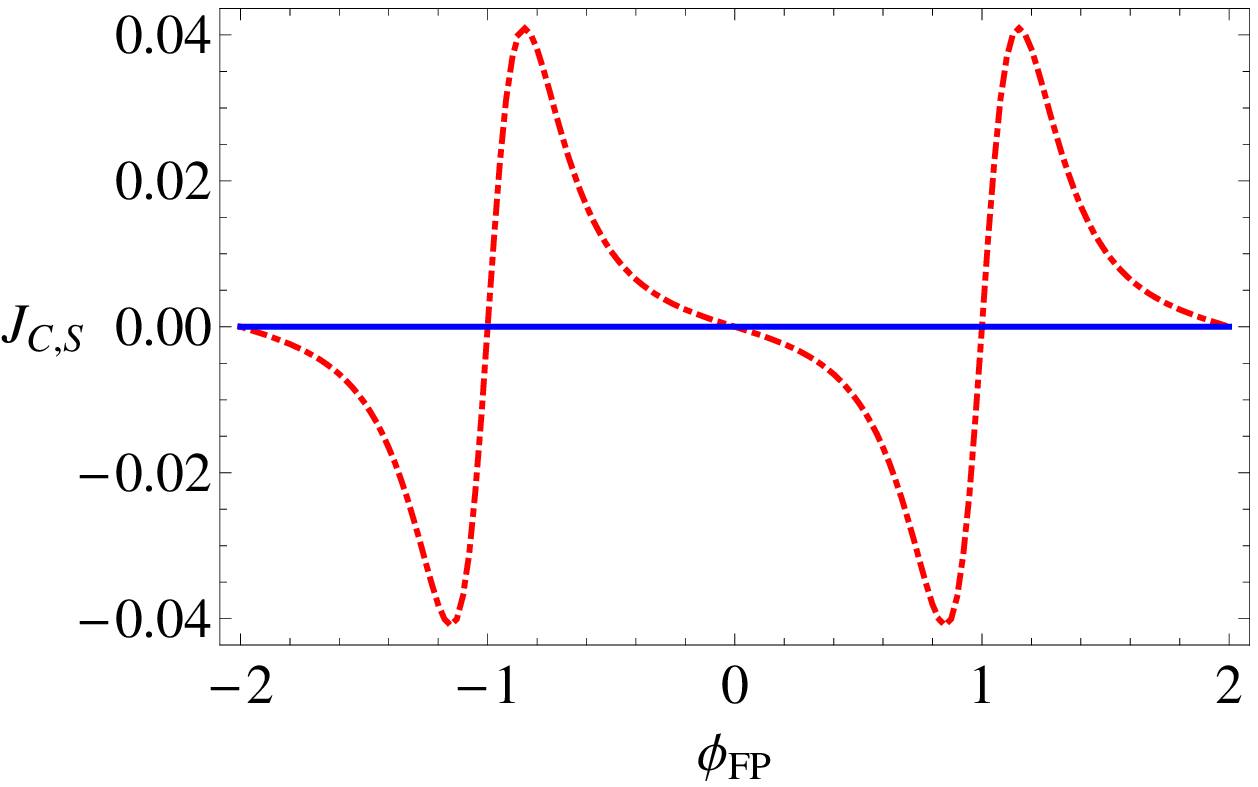}\\
\caption{The pumped charge (straight blue curve) and spin current (dotted red curve) in units of $e\omega_0/2\pi$ measured at the terminals 1 and 2 for the AMP (upper panel) and the PMP (lower panel). The parameters are $\gamma_{sp}^0=0.4$, $\gamma_{sf}^0=0.2$, $X_{1,2}^\omega=0.3$, $\varphi=\pi/2$, $\phi_{AB}=0.5$.}
\label{fig:ampl_pump_1}
\end{figure}

\begin{figure}
\centering
\includegraphics[scale=0.5]{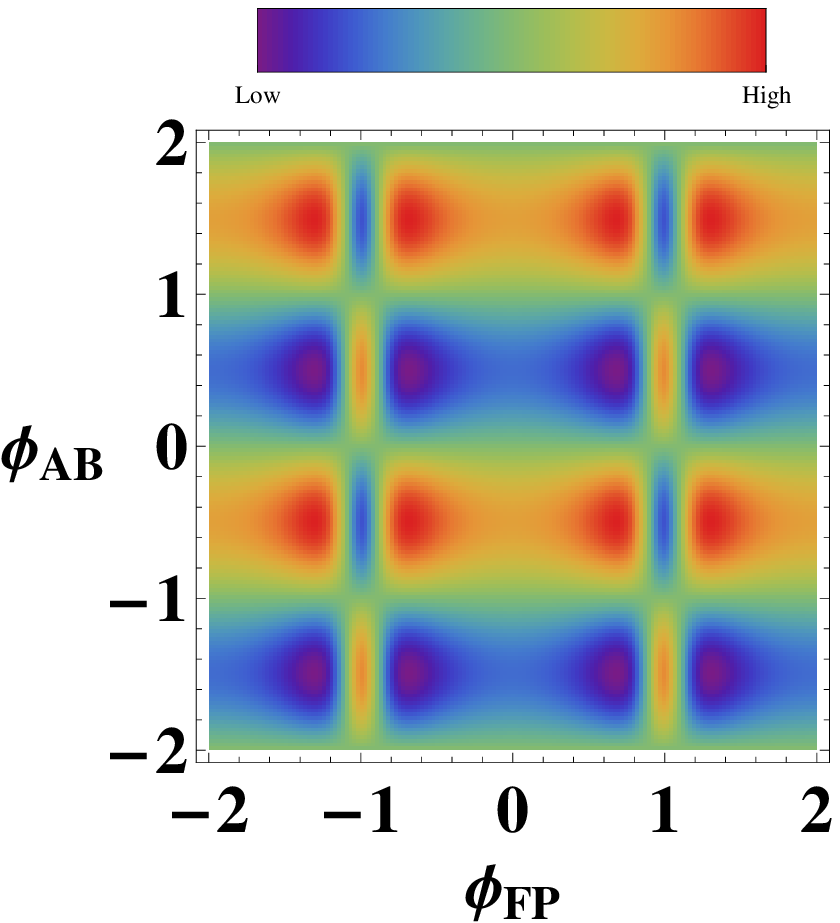}
\includegraphics[scale=0.5]{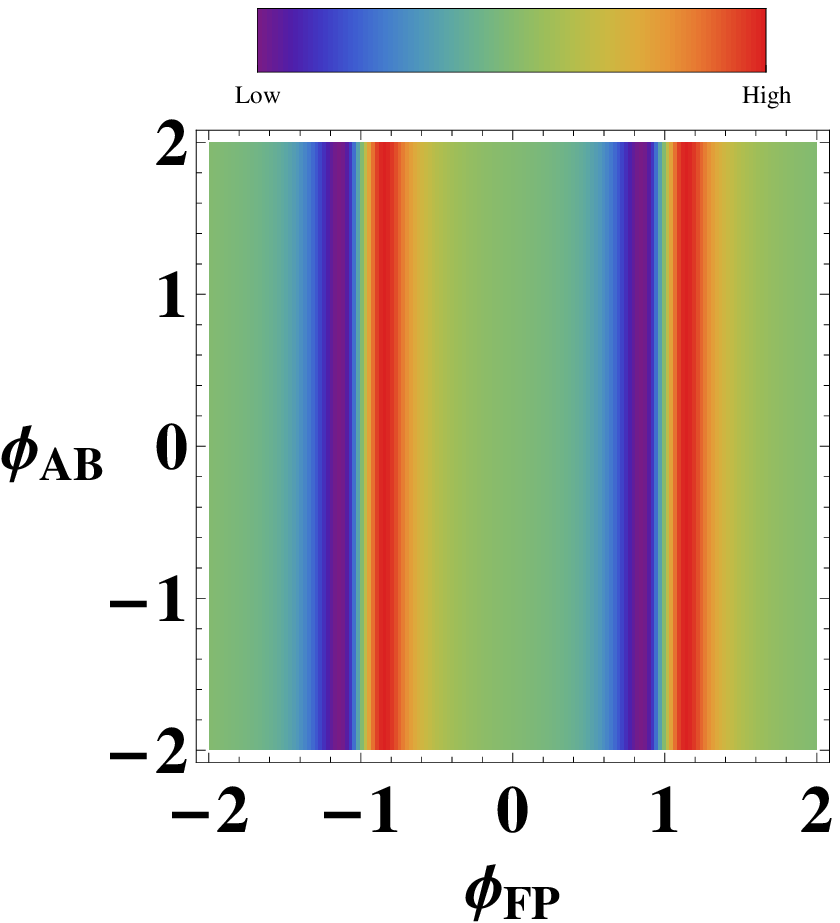}
\\
\caption{Contourplot of the spin current in units of $e\omega_0/2\pi$ for the AMP (left panel) and PMP (right panel). The other parameters are $\gamma_{sp}^0=0.4$, $\gamma_{sf}^0=0.2$, $X_{1,2}^{\omega}=0.3$, $\varphi=\pi/2$.}
\label{fig:count-pump}
\end{figure}

The result of the pumped charge and spin current for the AMP and PMP are shown in Fig.\ref{fig:ampl_pump_1} and Fig.\ref{fig:count-pump}.
In the case of AMP we always find a charge current $I_c^{pump}=\sum_{\sigma=\uparrow, \downarrow} (I_{R\sigma}^{pump}-I_{L \sigma}^{pump})$ along with a suppressed spin-current and a pumped charge $Q_p=\frac{2\pi}{\omega_0}I_c^{pump}\sim e$.
This peculiar behavior is shown in Fig.\ref{fig:ampl_pump_1} (upper panel) where the charge and spin currents are plotted as a function of the FP
phase while taking the AB flux half-integer. As  shown the spin current is almost nil except close to the zeros of the charge current. A quantization of the pumped charge $Q_p=2 e$ can be obtained at increasing the amplitude modulation from weak
to strong pumping regime. Let us note that the tunneling parameters, whose amplitude depends on the finite size effect of the constriction at the QPCs\cite{constr}, are chosen $\gamma^0_{sp}=0.4$ and $\gamma^0_{sp}=0.2$  which correspond  to
a spin conserving and spin-flipping transmission of 45$\%$ and 10$\%$, respectively, appropriate for real devices\cite{rothe_2010}.

In Fig.\ref{fig:count-pump} the spin-current is plotted in the plane $(\phi_{AB},\phi_{FP})$ for the AMP (left panel) and the PMP (right panel). In the case of AMP, zeros of the spin current, i.e. a pure charge current, are obtained for integer values of the AB flux, while zeros in both spin and charge currents occur at integer values of the FP phase in the PMP case.

Unlike the AMP, the results for the PMP show only a {\it pure spin} current for all the parameters range. It can be obtained with the gate voltages $V_{gt},V_{gb}$ and the expansion of the scattering matrix to lowest order in the tunneling amplitudes yields, $I_c^{pump}\simeq O(\gamma_{sp/sf}^4)$ and $I_s^{pump}\simeq -\frac{8e\omega_0}{2\pi} V^{\omega}_{gt}V^{\omega}_{gb} \gamma_{sp}^2 \sin \varphi \sin (\pi \phi_{FP})+O(\gamma_{sp/sf}^4)$.
For intermediate pumping frequencies the pumped spin charge $Q_s=\frac{2\pi}{\omega_0}I_s$ is not quantized and depends on the specific values of the external perturbations; it can be increased by increasing the parameters amplitude modulation. Universal quantized values for the pumped spin and charge currents, independent of the pumping cycle, can be found for interacting electrons in the asymptotic limits of slow or fast pumping frequencies, driven by the renormalization  of the couplings \cite{chamon_long}.

An important difference of our case compared  to the amplitude modulating electron pump studied in nanotubes or semiconducting rings is that there the spin-$\uparrow$ and spin-$\downarrow$ electrons are injected from the same electrode source and flow along the same channel. Here, the TI edge states are geometrically separated and the four terminal setup of Fig.1  allows for an independent control of the different electron species. The topological nature of the pump under investigation offers much richer physics and enables tunability of the type - charge or spin - as well as of the magnitude of the currents by {\it all electrical} means avoiding e.g. the complicated control of magnetization dynamics of magnetic proposals. Finally, our results  show that a quantum pumping provides a direct way to detect a TI phase by conventional
measurements of charge and spin  transport quantities. The locations of the zeros of the charge and spin currents have a distinct universal nature that can be easily detected in conventional charge measurements.

{\it Conclusions-} We proposed and analyzed  amplitude-modulating and  phase-modulating pumping in a double corner junction
in a TI insulator by {\it all-electrical} means as test bed
of  the nontrivial nature of the topological state.
The pure spin
current obtained via a phase modulating pumping is due to the exotic  response that can
only occur in a helical edge state and offers a fingerprint of the TI phase.
Measurements of spin transport in TI systems could be realized either via transport spectroscopy by direct measurements of spin polarization of emitted currents\cite{markus_spin_meas} or by spin-Hall effect and Kerr rotation measurements\cite{spintronics-ti-3}.

The authors acknowledge enlightening discussions with C.-H. Zhang, D. Bercioux, F. Dolcini and  E. Orignac.
\bibliographystyle{prsty}

\begin{thebibliography}{99}
\bibitem{note1} We defined: $\pi \phi_{FP}=\phi_T+\phi_B$ and $\pi \phi_{AB}=\phi_T-\phi_B$,
where $\phi_T=eV_{gt}L/\hbar v_F$ and $\phi_B=eV_{gb}L/\hbar v_F$.
\end{thebibliography}

\end{document}